\documentclass[journal]{IEEEtran}
\usepackage{amsmath,amsfonts}
\usepackage{algorithmic}
\usepackage{algorithm}
\usepackage{array}
\usepackage[caption=false,font=normalsize,labelfont=sf,textfont=sf]{subfig}
\usepackage{textcomp}
\usepackage{stfloats}
\usepackage{url}
\usepackage{verbatim}
\usepackage{graphicx}
\usepackage{multirow}
\usepackage{cite}
\usepackage{booktabs}
\usepackage{multirow}

\begin{document}

\title{A Plug-and-Play Untrained Neural Network for Full Waveform Inversion in Reconstructing Sound Speed Images of Ultrasound Computed Tomography}

\author{
\IEEEauthorblockN{Weicheng Yan, Qiude Zhang, Yun Wu, Zhaohui Liu, Liang Zhou, Mingyue Ding, Ming Yuchi, Wu Qiu}
\thanks{Weicheng Yan, Qiude Zhang, Yun Wu, Zhaohui Liu, Liang Zhou, Ming Yuchi, Ming Yuchi, and Wu Qiu are with Huazhong University of Science and Technology, Wuhan 430074, China.  Mingyue Ding is with Huazhong University of Science and Technology and Provincial-Ministerial Collaborative Major Science and Technology Infrastructure for High-end Biomedical Imaging, Wuhan 430074, China.} 
\thanks{Corresponding authors: Wu Qiu (wuqiu@hust.edu.cn) and Ming Yuchi
(m.yuchi@hust.edu.cn).}
}

\maketitle

\begin{abstract}
Ultrasound computed tomography (USCT), as an emerging technology, can provide multiple quantitative parametric images of human tissue, such as sound speed and attenuation images, distinguishing it from conventional B-mode (reflection) ultrasound imaging. Full waveform inversion (FWI) is acknowledged as a technique with the greatest potential for reconstructing high-resolution sound speed images in USCT. However, traditional FWI for sound speed image reconstruction suffers from high sensitivity to the initial model caused by its strong non-convex nonlinearity, resulting in poor performance when ultrasound signals are at high frequencies. This limitation significantly restricts the application of FWI in the  USCT imaging field. In this paper, we propose an untrained neural network (UNN) that can be integrated into the traditional iteration-based FWI framework as an implicit regularization prior. This integration allows for seamless deployment as a plug-and-play module within existing FWI algorithms or their variants. Notably, the proposed UNN method can be trained in an unsupervised fashion, a vital aspect in medical imaging where ground truth data is often unavailable. Evaluations of the numerical simulation and phantom experiment of the breast demonstrate that the proposed UNN improves the robustness of image reconstruction, reduces image artifacts, and achieves great image contrast. To the best of our knowledge, this study represents the first attempt to propose an implicit UNN for FWI in reconstructing sound speed images for USCT.  

\end{abstract}

\begin{IEEEkeywords}
Ultrasound computed tomography, Full waveform inversion, sound speed imaging, untrained neural network.
\end{IEEEkeywords}

\section{Introduction}
Ultrasound computed tomography (USCT), a state-of-the-art technology, exhibits significant potential in the fields of early breast cancer diagnosis, musculoskeletal imaging, and brain imaging \cite{park2023fast,duric2007detection,guasch2020full}. USCT utilizes a circular array of transducers to encircle the imaged object, capturing the ultrasound signal from all angles. This signal is then used to reconstruct three modal images: reflection image, sound speed image, and attenuation image. These images reflect the acoustic impedance, sound speed, and attenuation characteristics of human tissues, respectively, which are essential for clinical diagnosis. In this study, we focus on reconstructing high-quality sound speed images. \par
The most commonly employed sound speed reconstruction methods in practice are typically based on time-of-flight. These methods assume that sound waves propagate either along straight or bent lines, and subsequently calculate the distribution of sound speed in the region of interest by extracting the time of flight of the initial arriving wave. However, as these methods do not account for the physical phenomena of diffraction, reflection, and scattering of sound waves, they are characterized by high computational efficiency but low resolution, as well as a significant presence of artifacts. As a result, the utilization of sound speed images in clinical applications is severely restricted. The full-waveform inversion (FWI) technique has recently been studied extensively because of its potential to provide high-resolution sound speed images. FWI originated in the geophysical community to explore structures of the subsurface. It was subsequently introduced into the USCT field due to the similarity of the theory. However, the non-linear and non-convex nature of FWI makes it highly sensitive to the initial model. Failure to provide a sufficiently accurate initial model can lead the algorithm to converge toward a locally optimal solution, resulting in incorrect reconstruction.\par

\begin{figure}[!t]
\centering
\includegraphics[width=3in]{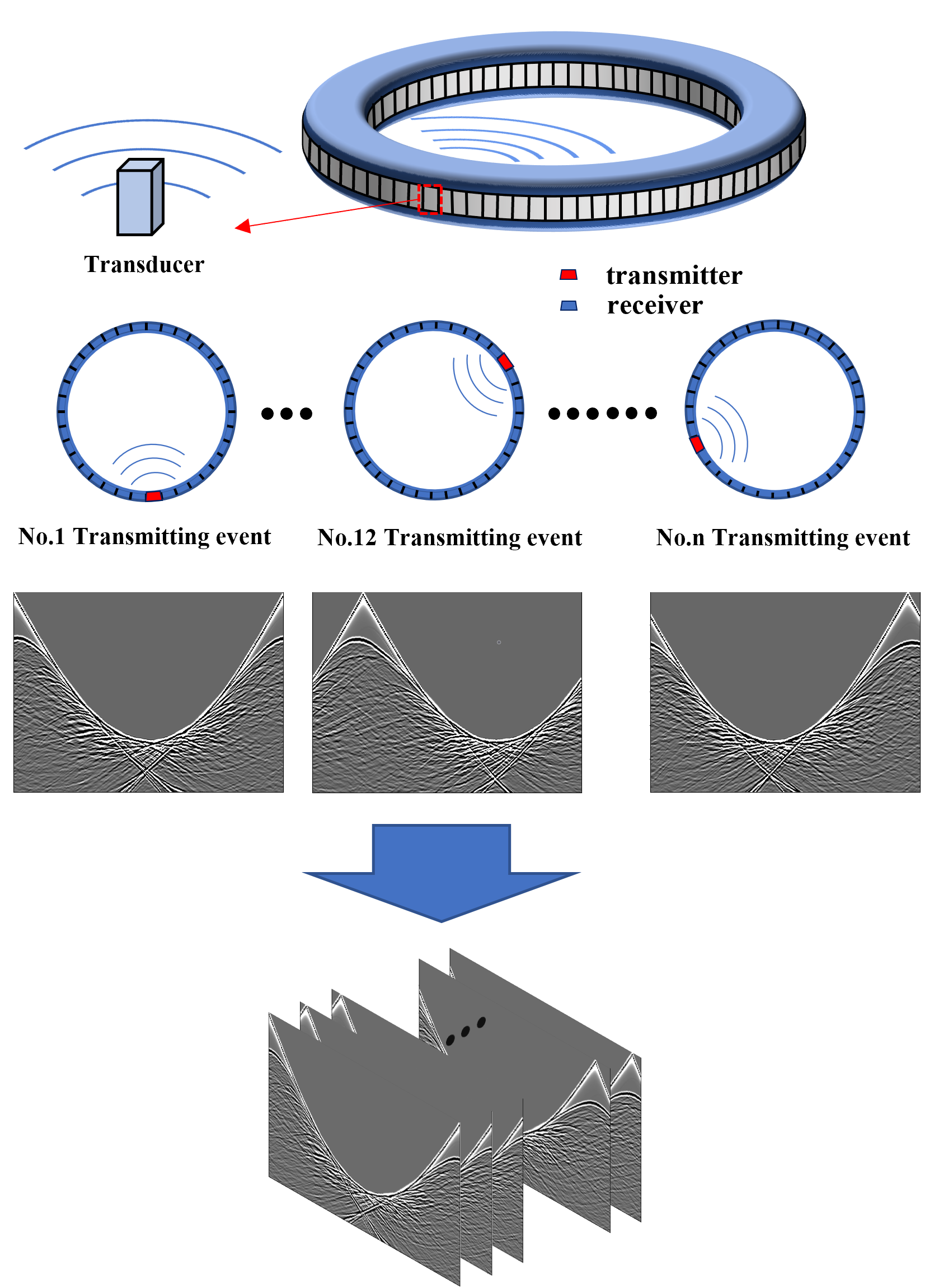}
\caption{Schematic diagram and workflow of the Ultrasound Computed Tomography system}
\label{fig_1}
\end{figure}

Many algorithms have been proposed to mitigate the sensitivity to initial model quality in the past decades. These methods can be broadly divided into three groups. The first group of techniques involves employing a different measure of the distance between the observed and PDE-simulated data, rather than the least squares misfit function. The second group of approaches is based on "extension strategies". This type of approach is very complex and time-consuming. It is not our purpose to give an extensive overview of these methods here, interested readers can study this paper \cite{operto2022full}. The third group of methods can be classified as deep neural network-based FWI. These methods employ either supervised or unsupervised learning techniques for sound speed reconstruction in FWI. \par
Nowadays, deep neural networks are powerful machine learning architectures with many applications. In the field of Full Waveform Inversion, several data-driven algorithms have been developed, including InversionNet\cite{wu2019inversionnet}, VelocityGAN\cite{zhang2020data}, and GAN-FWI\cite{mosser2020stochastic}. These algorithms employ various neural network architectures to learn from extensive datasets, enabling the network to accurately map signals to sound speed maps. However, the availability of large-scale paired datasets in both the geophysical and medical imaging domains is limited, posing challenges for practical applications. Another class of methods utilizes Physics-informed neural networks (PINNs), initially proposed for solving partial differential equations, which have later shown potential for addressing inverse problems. Khemraj successfully applied PINNs to the field of non-destructive testing, achieving accurate sound speed reconstruction of cracks\cite{shukla2020physics}. The final class of methods addresses the challenge of the limited availability of datasets by employing an unsupervised learning approach. These methods integrate the network into the algorithmic flow of FWI, such as UPFWI, NNFWI, DNN-FWI\cite{jin2021unsupervised,zhu2022integrating,he2021reparameterized}, and so on. These approaches assume that the network structure itself can provide regularization to FWI and mitigate its sensitivity to the initial model. Untrained neural networks (UNN) have demonstrated their effectiveness as prior knowledge in solving inverse problems by fitting a neural network model to the input\cite{qayyum2022untrained}. This idea was first noted in the deep image prior\cite{ulyanov2018deep}.\par
UNN was first proposed to handle image post-processing tasks such as image denoising, inpainting, and super-resolution \cite{ulyanov2018deep}. Due to its excellent performance in inverse imaging problems, it has thus been introduced to PET, MRI, and CT reconstruction problems\cite{baguer2020computed,gong2018pet,yoo2021time}. Inspired by the preceding works, we apply the UNN to the field of USCT sound speed imaging. To the best of our knowledge, this is the first study of incorporating the UNN module to sound speed reconstruction in USCT. In this study, we propose a plug-and-play UNN module that can be easily embedded into the framework of existing FWI or its variants algorithms. Our method introduces the regularization effect, which is caused by the network structure, into the reconstruction of sound speed in USCT. In detail, we utilize a generator network to produce a sound speed image. This image is then incorporated into the iterative process of FWI as the sound speed map. Unlike traditional methods, the proposed approach no longer directly updates the sound speed image. Instead, it indirectly updates the sound speed image by adjusting the network parameters of the generator. We evaluate the effectiveness of our proposed approach through the numerical simulation and phantom experiment of the breast. Our results demonstrate that the proposed method significantly reduces the sensitivity of FWI to the initial model and substantially improves the stability of FWI in USCT sound speed reconstruction, especially at high frequencies.\par
The structure of the article is organized as follows. In section II, we provide a concise overview of the background of USCT and the principles of FWI technology for sound speed reconstruction. Additionally, it presents the UNN module and provides a detailed explanation of the proposed method. Next, we provide the numerical simulation and phantom experiment implementation details and results in Sections III and IV, respectively. Section V discusses the robustness, potential, and challenges of the proposed approach. Lastly, conclusions are presented in Section VI.

\section{Methods}
In this section, we first describe how sound speed images are reconstructed with FWI in USCT, then briefly describe the contents of untrained neural networks prior, and finally present our method of embedding UNN in FWI to reconstruct sound speed in USCT.
\subsection{Full Waveform Inversion in USCT}
Full waveform inversion is a highly promising approach for the inverse problem of sound speed imaging in the USCT. Mathematically, FWI can be regarded as an optimization problem subject to the constraint of a partial differential equation (PDE). Under the constraint of the acoustic wave equation, sound speed is determined by minimizing the misfit between the observed and PDE-simulated data\cite{virieux2009overview}, i.e.,
\begin{equation}
\label{deqn_ex1a}
\begin{array}{cc}
\hat{c}=\underset{c}{\arg \min }\underset{i}{\overset{N_{s}}{\sum}}\underset{j}{\overset{N_{r}}{\sum}}\|u_{obs}(i,j)-Ru(c)\|^{2},\\
\text { s.t. } \frac{\partial^2u}{\partial t^2} - c^2\Delta u = f(x,t),
\end{array}
\end{equation}
where $c$ is the sound speed of the model, $f$ is the source term which is a function of spatial and temporal, $u_{obs}$ is the observed data, and $u(c)$ is the PDE-simulated data determined by the sound speed $c$. The sampling operator $R$ indicates the location of receiver transducers in the spatial domain and $N_{s}$, $N_{r}$ are the number of transmitting events and the number of receiver transducers, respectively. In USCT, the position information of the ring array is embedded in the sampling operator $R$. $u_{obs}$ consists of $N_{r}$ transmitting events, in each of them only one array element serves as the transmitter and the rest of the arrays serve as receivers, and the full-view acquisition of ultrasound signals is accomplished by switching the position of the transmitter, as illustrated in Fig. 1.\par
For simplicity, we replace the acoustic wave equation with the forward operator $\mathcal{W}(\cdot): \mathbf{I}  \in \mathbb{R}^{3} \mapsto \mathbf{D} \in \mathbb{R}^{3} \times [0, T)$, which denotes the mapping from the physical domain ($\mathbf{I}$) to the data domain ($\mathbf{D}$). So we can reformulate Eq. 1 as
\begin{equation}
\label{deqn_ex1a}
\hat{c}=\underset{c}{\arg \min }\underset{i}{\overset{N_{s}}{\sum}}\underset{j}{\overset{N_{r}}{\sum}}\|u_{obs}(i,j)-R\mathcal{W}(c)\|^{2}.
\end{equation}
The workflow of FWI is shown in Fig. 2. However, the minimization of Eq. 1 is a nonlinear and nonconvex optimization problem, the constraint of the PDE makes FWI severely ill-posed. If the initial sound speed model is too distant from the global minimum, FWI will converge towards a local minimum that may provide a non-physiological meaning result. This phenomenon is also known as cycle-skipping. In this article, we embed the UNN framework into FWI for regularising such an ill-posed problem thus avoiding cycle-skipping.

\begin{figure}[!t]
\centering
\includegraphics[width=3in]{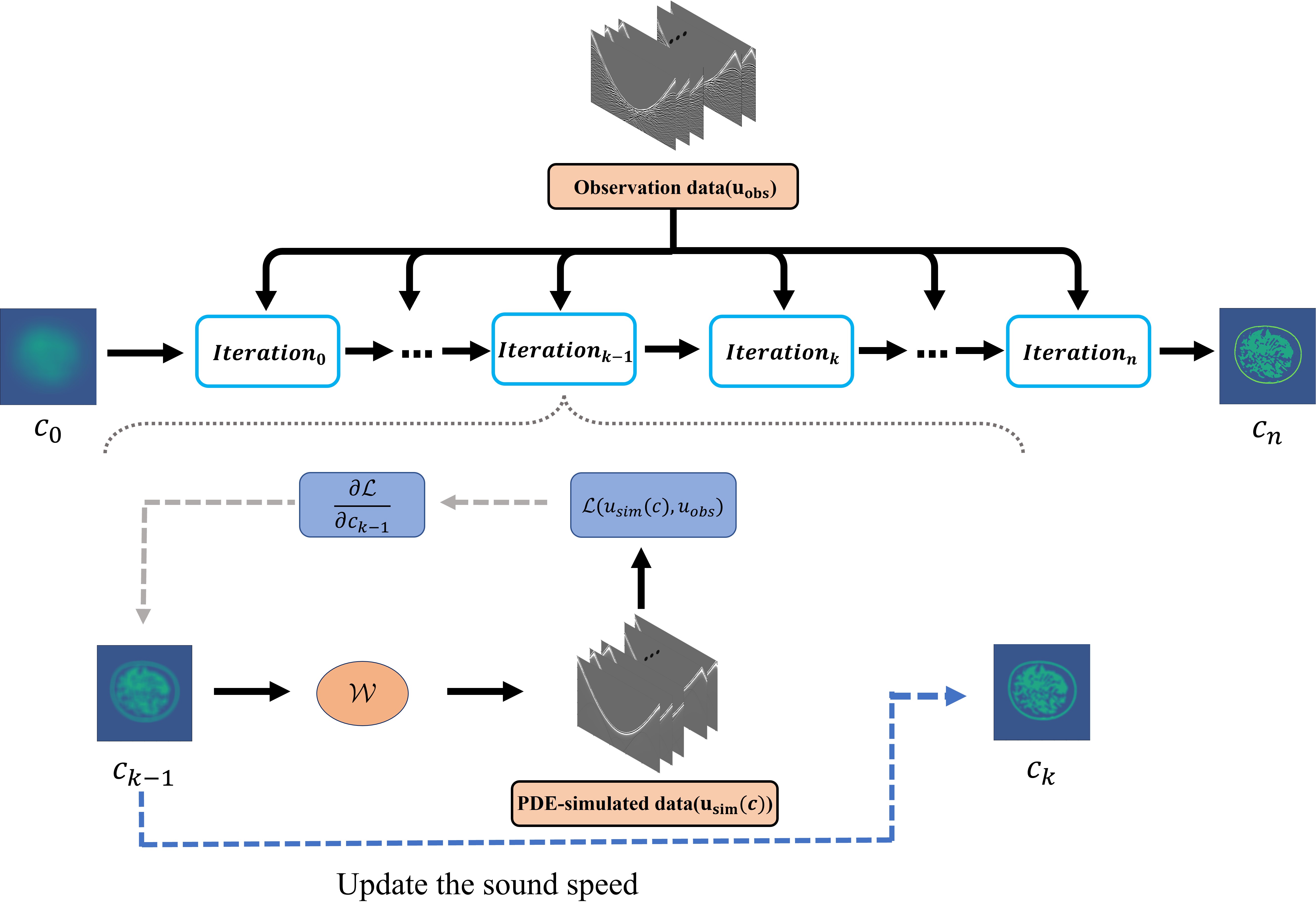}
\caption{Workflow of Full Waveform Inversion in Reconstructing Sound Speed Images of Ultrasound Computed Tomography}
\label{fig_2}
\end{figure}

\subsection{Untrained Neural Networks Prior}
The UNN uses the structure of the neural network as a prior to regularising the inverse problem. In the UNN framework, instead of using explicit regularization terms, the reparametrization of the unknown variable is used for implicit regularization. More specifically, the unknown, $x$, is generated by a neural network, $x$ = $\mathcal{N}_{\theta}(z)$, where $\mathcal{N}(\cdot)$ indicates the structure of the neural network, $\theta$ stands for the network’s parameters to be learned, and $z$ is a fixed random noise. Thus the UNN framework can be formulated as
\begin{equation}
\label{deqn_ex1a}
\hat{\theta} = \underset{\theta}{\arg \min}\mathcal{F}(\mathcal{N}_{\theta}(z)), \quad \hat{x} = \mathcal{N}_{\hat{\theta}}(z),
\end{equation}
where $\hat{x}$ is the final desired solution of the inverse problem, and $\mathcal{F}(\cdot)$ indicates any misfit function. Through a deep generative network, unknown variables are obtained indirectly by training the network parameters $\theta$, this process is the so-called reparametrization. From another point of view, the generator networks achieve a sparse representation of unknown variables. The success of UNN may be ascribed to the implicit representation of the network architecture which can be seen as an implicit optimized regularizer\cite{cheng2019bayesian}.\par

In the range of the inverse problem, the unknown variables obtained from the output of the neural network can be readily employed as inputs for classical model-driven algorithms like FWI. Hence, the UNN framework trains the network parameters in a purely unsupervised way, without any pre-training or labeling data constraints during the training process. This is important for sound speed reconstruction in USCT because of the lack of high-quality clinical image data sets. So embedding the UNN framework in the sound speed reconstruction process is straightforward and easy to implement.

\begin{figure}[!t]
\centering
\subfloat[\label{fig:a}]{\includegraphics[width=3.5in]{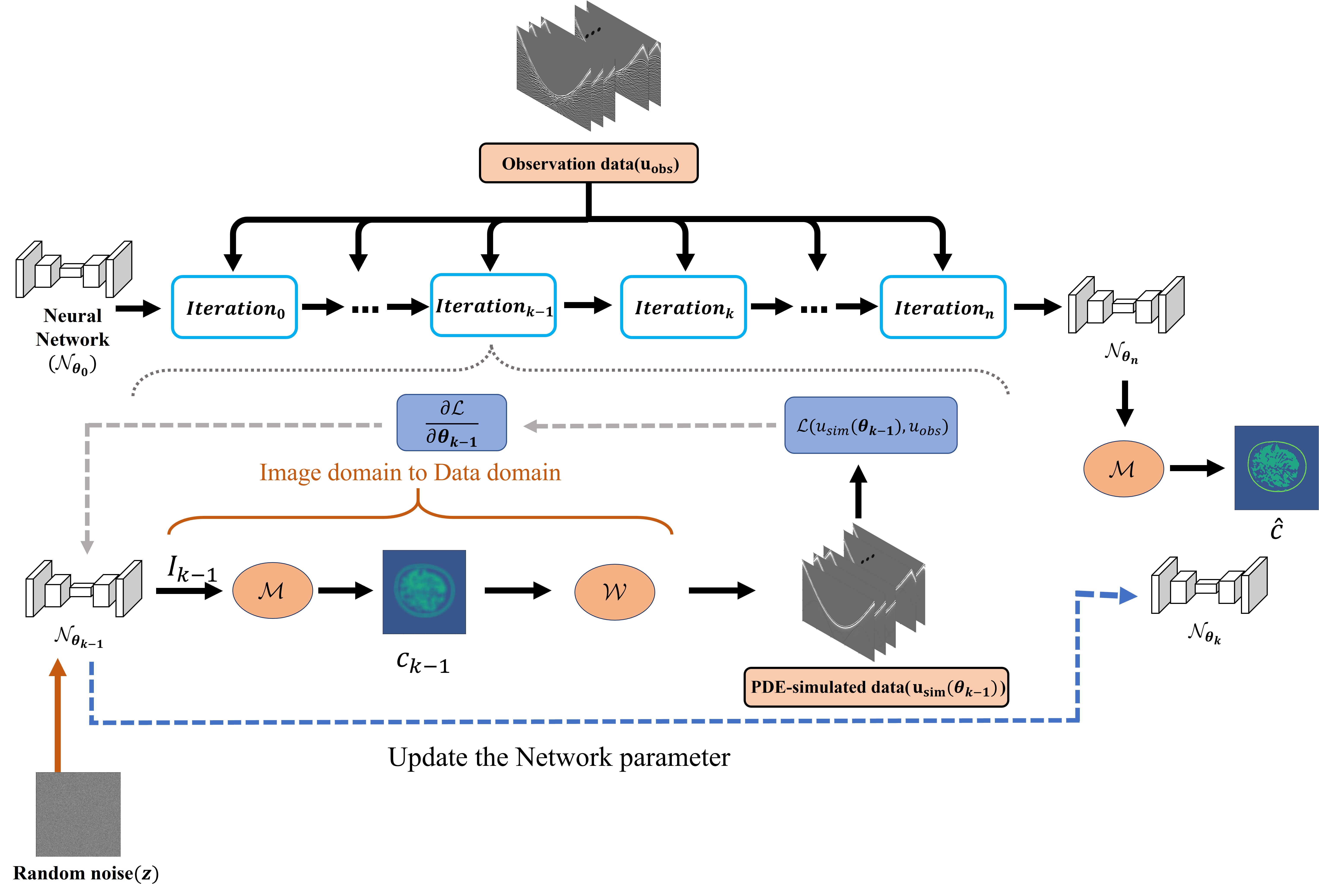}}
\\
\subfloat[\label{fig:b}]{\includegraphics[width=3.5in]{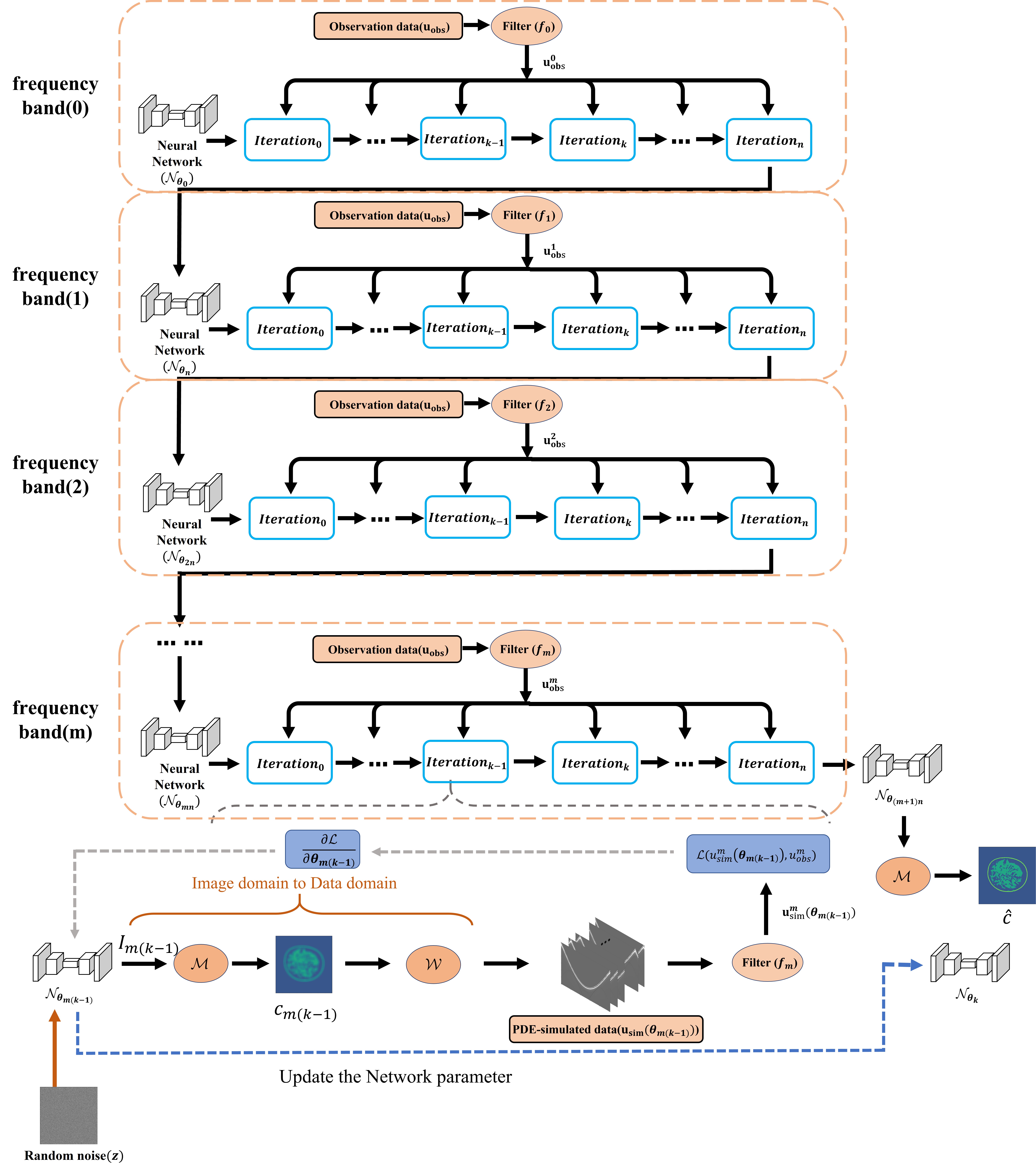}}
\caption{The flowchart of the proposed method, $I_{i}$ is the output image of the network, $c_{i}$ is the sound speed map, $u(\theta_{i})$ is the predicted ultrasound signals obtained from the sound speed map. (a) is the framework of UNN-FWI, and (b) is the framework of UNN-MFWI.}
\label{fig_3}
\end{figure}

\subsection{Proposed Framework}
To avoid the cycle-skipping problem in sound speed reconstruction, we propose a new framework, full waveform inversion integrated with untrained neural networks (UNN-FWI). Before introducing UNN-FWI, we return to Eq. 2 for a detailed explanation. Since the forward operator $\mathcal{W}$ generally has no analytic solution, numerical methods such as finite difference method (FDM), finite element method, and so on are usually used to find the solution. In this article, the numerical solution of $\mathcal{W}$ is obtained using FDM due to its simplicity and efficiency. \par

Recalling the UNN framework discussed in the previous subsection, the main idea is to use a neural network to generate a sound speed image. Since the output of the network has no physical meaning, we need a transformation function $\mathcal{M}(\cdot)$ to convert the network output from the image domain to the physical domain, which is necessary for the forward operator. The transformation function must adhere to the following three rules: 1. The output of the network should not contain negative values as sound speed cannot be negative. 2. Given that the physical parameters of the forward operator (e.g., spacing step and timing step) are determined, the speed of sound must fall within a specified range to guarantee the stability and accuracy of the calculation. Thus, the output of the transformation operator needs to be within this range. 3. The transformation operator should retain the image domain information contained in the network's output. In this study, we employ the sigmoid layer as the output layer of the network to restrict the image values within the range of 0 and 1. Furthermore, assuming that the speed of sound is confined to a specific range denoted as [$c_{min}$, $c_{max}$], we can define the transformation function as follows:
\begin{equation}
\label{deqn_ex1a}
\mathcal{M}_{1}(x) = (c_{max} - c_{min})x + c_{min},
\end{equation}
\begin{equation}
\label{deqn_ex1a}
\mathcal{M}_{2}(x) = (c_{max} - c_{min})^x + c_{min}.
\end{equation}
$\mathcal{M}_{1}$ represents a linear function, while $\mathcal{M}_{2}$ corresponds to an exponential function. Although numerous other functions could fulfill the conditions of the transformation operator, we will limit our examples to these two. We will now introduce the UNN-FWI architecture and its variant UNN-MFWI. The latter one will show the compatibility of the UNN as a plug-and-play module when combined with other strategies used in the field of FWI.\par

\subsubsection{UNN-FWI}
With the implementation of transformation operators and neural networks, the minimization problem (3) can be reformulated as:
\begin{equation}
\label{deqn_ex1a}
\begin{array}{cc}
\hat{\theta} = \underset{\theta}{\arg \min }\underset{i}{\overset{N_{s}}{\sum}}\underset{j}{\overset{N_{r}}{\sum}}\|\mathcal{D}_{\theta}\|^{2},\\\\
\mathcal{D}_{\theta} = (u_{obs}(i,j)-R\mathcal{W}(\mathcal{M}(\mathcal{N}_{\theta}(z))), \\\\
\hat{c} = \mathcal{M}(\mathcal{N}_{\hat{\theta}}(z)).
\end{array}
\end{equation}
This is what we call the UNN-FWI framework and Fig. 3(a) illustrates its workflow. $\mathcal{L}$ represents the loss function, which quantifies the discrepancy between predicted and observed values, and we use the least-squares norm as the chosen loss function in this article.\par
\subsubsection{UNN-MFWI}
Multi-band strategy is often employed in FWI to enhance the robustness of the reconstruction, particularly in high-frequency scenarios\cite{pratt1999seismic}. The main idea behind this approach is to initially minimize the difference between the low-frequency predicted and observed signals, and subsequently, progress in reducing the discrepancy step by step as the frequencies increase. So we can rewrite Eq. 6 as:
\begin{equation}
\label{deqn_ex1a}
\begin{array}{cc}
\hat{\theta} = \underset{\theta}{\arg \min }\underset{i}{\overset{N_{s}}{\sum}}\underset{j}{\overset{N_{r}}{\sum}}\|f \ast \mathcal{D}_{\theta}\|^{2},\\\\
\mathcal{D}_{\theta} = (u_{obs}(i,j)-R\mathcal{W}(\mathcal{M}(\mathcal{N}_{\theta}(z))), \\\\
\hat{c} = \mathcal{M}(\mathcal{N}_{\hat{\theta}}(z)).
\end{array}
\end{equation}
$f$ is a low-pass filter with a cutoff frequency that progressively increases as the number of iterations increases and Fig. 3(b) illustrates its workflow.\par

Next, we will elaborate on the process of training the network, specifically addressing the flow of gradients from the loss function to the network parameters. From the law of chains, we know that:
\begin{equation}
\label{deqn_ex1a}
\frac{\partial \mathcal{L}}{\partial \theta} = \frac{\partial \mathcal{L}}{\partial c} \frac{\partial c}{\partial I} \frac{\partial I}{\partial \theta}.
\end{equation}
The first term on the right-hand side of Eq. 7 is the gradient of sound speed. In the classical FWI framework, in order to calculate the gradient, the adjoint method is utilized, which involves computing the zero-lag cross-correlation between the adjoint wavefield and the time derivative of the incident wavefield. The second term is related to the transformation operator and the third term can be calculated automatically by back-propagation. Hence, the gradient can be calculated and network parameters updated through the utilization of the chain rule, albeit with some complexity. Recently, Richardson proposed utilizing recurrent neural networks for modeling wave equation and incorporating automatic differentiation in deep learning framework to compute the gradient\cite{richardson2018seismic}. The method of full waveform inversion with automatic differentiation has gained considerable attention in recent years due to its mathematical equivalence to the adjoint method and its compatibility with other neural networks \cite{zhu2021general,zhou2022efficient}. In this article, we employ the automatic differentiation framework for computing the gradient instead of utilizing the complicated adjoint method. We consider the forward operator $\mathcal{W}$ as a fixed-parameter neural network. Consequently, the gradient can be effortlessly computed using automatic differentiation techniques, and the generator network training can utilize any standard optimization algorithms.

\section{Experiments}
This section presents a detailed application of the proposed method to Ultrasound Computerized Tomography and investigates its advantages through the numerical simulation and phantom experiment performed on the breast. This section first presents the network architecture employed, followed by an explanation of the numerical and phantom study setup and the specific application of the finite difference method. Lastly, the evaluation criteria and baseline method are described.
\subsection{Network Architecture}
 In the original UNN prior paper\cite{ulyanov2018deep}, the best results can be achieved by choosing different network structures for different tasks. For instance, the autoencoder exhibits remarkable performance in inpainting tasks, while the U-Net delivers optimal results in denoising tasks. The regularization effect of the UNN is primarily imposed by the network structure, which confines the solution space of the generated sound speed image. This confinement enables the optimization process of FWI to converge more easily and effectively to the global optimal solution. In this study, we use U-Net with depth 5 and full skip connections architecture. Full skip connections are used to make network training more stable. The U-Net model has an input channel of 1, as the input random noise z does not necessarily need to be an RGB image. The output layer is the sigmoid layer, as previously mentioned. Each layer has a different number of filters: (64, 128, 256, 512, 1024, 512, 256, 128, 64). This design enables a more robust regularization effect. We use LeakyReLU as the activation function of the hidden layer and the Bi-linear operator as the up-sampling operator. In the convolution layer, the convolution kernel size is 3×3, the stride is 1×1 and the padding is zero. The neural network architecture is illustrated in the TABLE. I.

\subsection{Numerical Setting and Implementation Details}
In this paper, we perform the numerical simulation and phantom experiment of the breast, to investigate the performance of the proposed method. In the numerical study, the computational domain was defined as a square with a side length of 0.25 m in space and a duration of 0.2e-3 s in time. A ring array of transducers, similar to those used in USCT, was employed, consisting of 256 transducers with a radius of 0.12 m. Among these transducers, 64 uniformly distributed transducers were designated as emitters, and for each transmitting event, one transducer was designated as the transmitter and the remaining transducers served as receivers. In the phantom experiment, the computational domain was defined as a square with a side length of 0.23 m in space and a duration of 0.16e-3 s in time. To scan the two-dimensional tomographic surface of the phantom, a ring-array ultrasound transducer with a diameter of 0.22 m, composed of 512 individual transducers, was employed. Each transducer has a center frequency of about 500 kHz.\par 
In the study of the numerical simulation, the model was developed using a database of MRI segmentation images \cite{lou2017generation}. We subsequently assign different acoustic properties to the images depending on the tissue. The background was assumed to be water with a sound speed set at 1480 m/s. The sound speeds for the skin, fat, gland, blood vessels, and tumor were set at 1650 m/s, 1470-1490 m/s (following a Gaussian distribution), 1580 m/s, 1600 m/s, and 1630 m/s, respectively, as shown in the top left corner of Fig. 4. We perform reconstruction experiments using a Ricker source at 500kHz and 800kHz frequencies as the emitted sound source to demonstrate the effectiveness of the proposed method at different frequencies. The simulated observed signals are generated by the 2D-FDM at a frequency of 500 kHz, with a spatial grid size of Ns = 800 and a temporal grid size of Nt = 6000. To avoid the issue of 'inverse crime', the reconstruction process utilizes a smaller spatial grid size of Ns = 592 and a temporal grid size of Nt = 3000. Similarly, at a frequency of 800 kHz, the observed signals are generated with a spatial grid size of Ns = 1000 and a temporal grid size of Nt = 8000, while the inverse process employs a spatial grid size of Ns = 800 and a temporal grid size of Nt = 4400.\par
In the study of the phantom experiment, the shape of the phantom used was elliptical, as shown in Fig.5. The background sound speed within the breast phantom was set at 1540 m/s. To assess the resolving capability of the sound speed imaging, a cylindrical object with a diameter of 3 mm and a sound speed of 1580 m/s was strategically inserted vertically within the phantom. We use a spatial grid size of Ns = 896 to ensure the computational accuracy of the finite difference method. The sampling rate of the observed signal is 25 MHz, so we have a temporal grid size of Nt = 4000. For the observation signals, due to the simpler structure of the imaging object, we only used 256 transmit events, each event consisting of 512 received signals. In practical implementation, not all received signals will be utilized, because our transmitting transducers cannot be used as point sources as in numerical simulations. Consequently, we exclude 64 signals closest to the transmitting transducer due to their significant error.\par
Next, some implementation details are presented. A Gaussian distribution with mean 0 and variance 0.1 is sampled for the input noise image $z$. As previously mentioned, we employ the Deepwave \cite{deepwave}, a Python package built on PyTorch, to establish a fixed-parameter recurrent neural network. This network serves as the forward operator $\mathcal{W}$ and is employed to replace the finite difference method. We adopt the Adam optimizer to optimize the network parameters with a learning rate lr = 1e-4 and the maximum number of iterations $n_{iter}$ = 5000. In this study, all the experiments are implemented in Intel i7-11700 CPU @ 3.60 GHz CPU and NVIDIA RTX4090 GPU. Pytorch 2.1.0 on Python 3.9.7 is used to implement our generative network.\par

\begin{table*}[ht]
    \centering
    \caption {Network Architecture}
    \label{tab:chap:table_1}
    \begin{tabular}[c]{cccccc}
	\toprule
	{Operation Layer} & {Number of Filters} & {Size of Each Filter } & {Stride} & {Skipping Connection} & {Size of output image} \\
	\midrule
	Input image $z$ &  &  &  &  & 1 × W × W \\
	2×(Conv+BN+ReLU) & 64 & 3 × 3 × 64 & 1 × 1 &  & 64 × W × W (encode layer1) \\
	Maxpool &  &  & 2 × 2  &  & 64 × W/2 × W/2 \\
	2×(Conv+BN+ReLU) & 128 & 3 × 3 × 128 & 1 × 1 &  & 128 × W/2 × W/2 (encode layer2)  \\
	Maxpool &  &  & 2 × 2  &  & 128 × W/4 × W/4 \\        
 	2×(Conv+BN+ReLU) & 256 & 3 × 3 × 256 & 1 × 1 &  & 256 × W/4 × W/4 (encode layer3)  \\
	Maxpool &  &  & 2 × 2  &  & 256 × W/8 × W/8 \\   
 	2×(Conv+BN+ReLU) & 512 & 3 × 3 × 512 & 1 × 1 &  & 512 × W/8 × W/8 (encode layer4) \\
	Maxpool &  &  & 2 × 2  &  & 512 × W/16 × W/16 \\   
        2×(Conv+BN+ReLU) & 1024 & 3 × 3 × 1024 & 1 × 1 &  & 1024 × W/16 × W/16  \\
        Up-conv + concat &  &  & 2 × 2 & encode layer4 & 1024 × W/8 × W/8  \\
        2×(Conv+BN+ReLU) & 512 & 3 × 3 × 512 & 1 × 1 &  & 512 × W/8 × W/8  \\
        Up-conv + concat &  &  & 2 × 2 & encode layer3 &  512 × W/4 × W/4  \\
        2×(Conv+BN+ReLU) & 256 & 3 × 3 × 256 & 1 × 1 &  & 256 × W/4 × W/4  \\
        Up-conv + concat &  &  & 2 × 2 & encode layer2 &  256 × W/2 × W/2  \\
        2×(Conv+BN+ReLU) & 128 & 3 × 3 × 128 & 1 × 1 &  & 128 × W/2 × W/2  \\
        Up-conv + concat &  &  & 2 × 2 & encode layer1 &  128 × W × W  \\
        2×(Conv+BN+ReLU) & 64 & 3 × 3 × 64 & 1 × 1 &  & 64 × W × W  \\
        Conv & 1 & 1 × 1 × 1 &  1 × 1 &  & 1 × W × W \\
        Sigmoid          &    &   &   &  &  1 × W × W  \\
	\bottomrule
    \end{tabular}
\end{table*}

\subsection{Baseline Methods and Evaluation Metric}
The classical least squares-based full waveform inversion (L2-FWI) method is employed as the baseline method in this study. To more comprehensively investigate the multi-band strategy, we incorporate it into L2-FWI, resulting in L2-MFWI for comparison. While various FWI variant algorithms exist, which are based on different loss functions like AWI (based on the Wiener filter) \cite{warner2016adaptive} and OT-FWI (based on Optimal Transport Distance) \cite{engquist2022optimal}, all of these methods can be seamlessly integrated into the UNN-FWI architecture. Therefore, for simplicity, only the least squares loss function is employed as the baseline method. Additionally, the same initial model was employed for both L2-FWI and L2-MFWI methods. In the breast experiment, a homogeneous sound field with a speed of 1480 m/s was utilized, while in the brain experiment, a homogeneous sound field with a speed of 1500 m/s was employed.\par
We employ three quantitative evaluation metrics: peak signal-to-noise ratio (PSNR), structural similarity index (SSIM) \cite{hore2010image}, and root mean square error (RMSE). Higher scores of PSNR and SSIM indicate better reconstruction quality. A lower value of RMSE represents a smaller error in the speed of sound of the reconstructed image. With the ground truth $x$ and the reconstructed image $x^*$, RMSE, PSNR ,and SSIM are given by
\begin{equation}
\label{deqn_ex1a}
\begin{array}{cc}
RMSE  = \sqrt{\frac 1 {mn} \sum_{i=0}^{m-1} \sum_{j=0}^{n-1} (x(i,j)-x^*(i,j))^2}, \\\\
PSNR = 10 log_{10}(\frac {MAX_I} {MSE})^2, \\\\
SSIM = \frac{(2\mu_{x^*}\mu_x+c_1)(2\sigma_{x^*x}+c_2)}{(\mu_{x^*}^2 + \mu_{x}^2 + c1)(\sigma_{x^*}^2 + \sigma_{x}^2) + c2}.
\end{array}
\end{equation}
$MAX_I$ is the maximum possible pixel value of the image, $\mu_{x^*}$ and $\mu_x$ are the expectations of the reconstructed image and the ground truth, $\sigma_{x^*}^2$ and $\sigma_{x}^2$ are the variances, and $\sigma_{x^*x}$ is the covariance. In addition, c1 and c2 are stabilization parameters. In this study, since the reconstructed images are sound speed values, we use the network outputs image without transformation functions when calculating the PSNR and SSIM metrics, and when calculating the MSE, we use the final reconstructed sound speed images for the sake of accuracy.

\begin{figure*}[!t]
\centering
\subfloat[\label{fig:a}500KHz of Breast]{\includegraphics[width=6.5in]{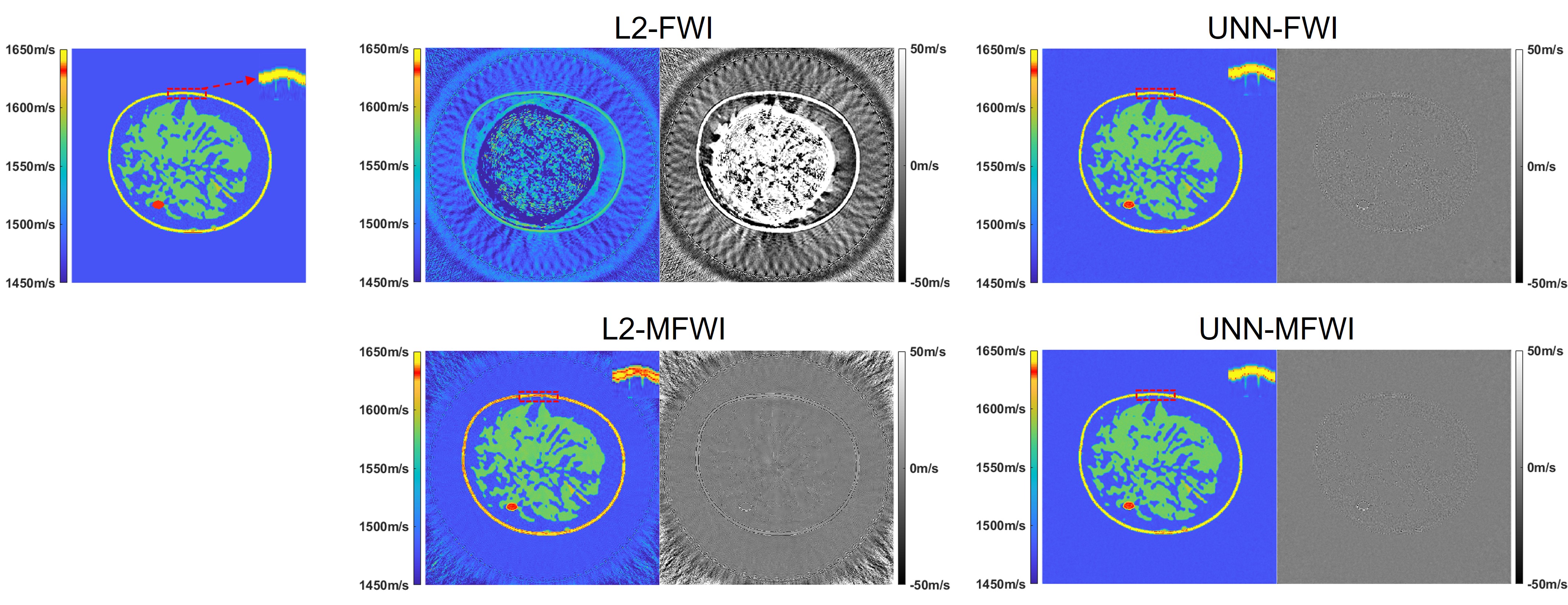}}\\
\subfloat[\label{fig:b}800KHz of Breast]{\includegraphics[width=6.5in]{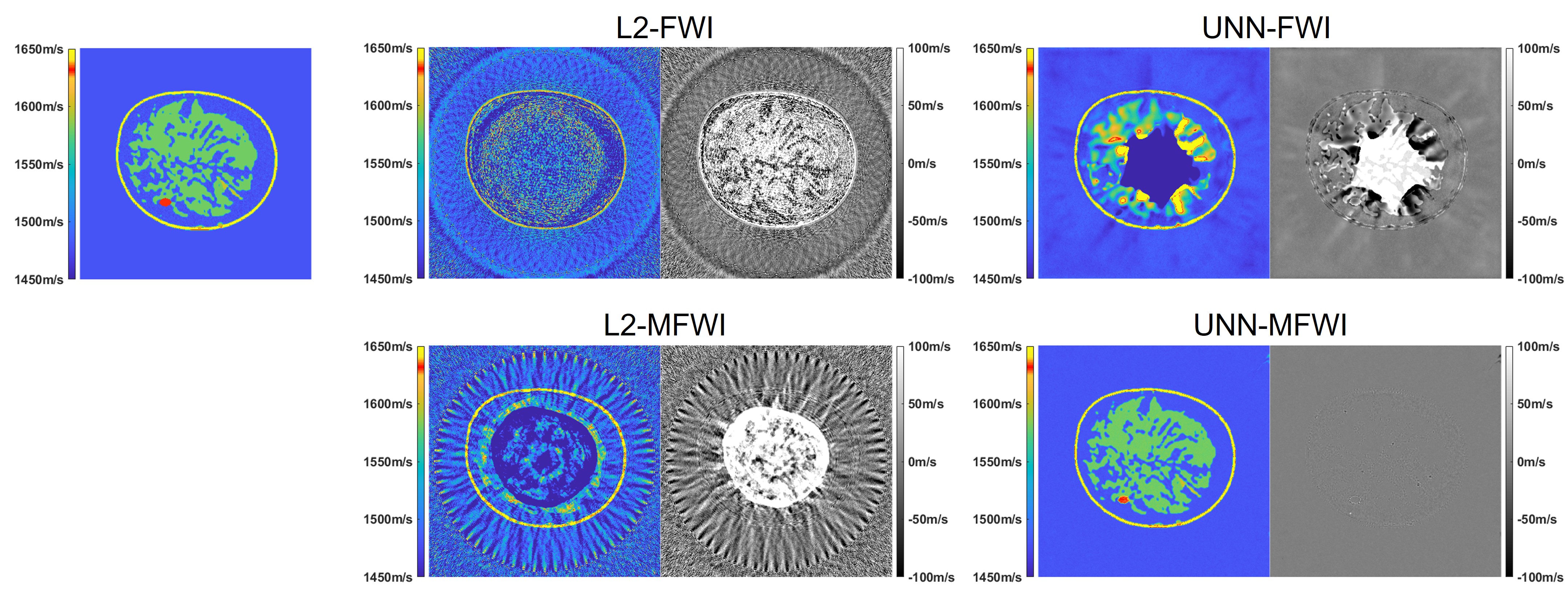}}
\caption{Reconstructed sound speed images and error images obtained by different methods for numerical experiments on the breast. (a) and (b) are breast experiments at 500K and 800K center frequencies, respectively. In each subfigure, the image in the upper left corner is the ground truth, in the results of the different methods the left column is the sound speed image the right column is the error image.}
\label{fig_4}
\end{figure*}

\begin{figure*}[!t]
\centering
\includegraphics[width=6in]{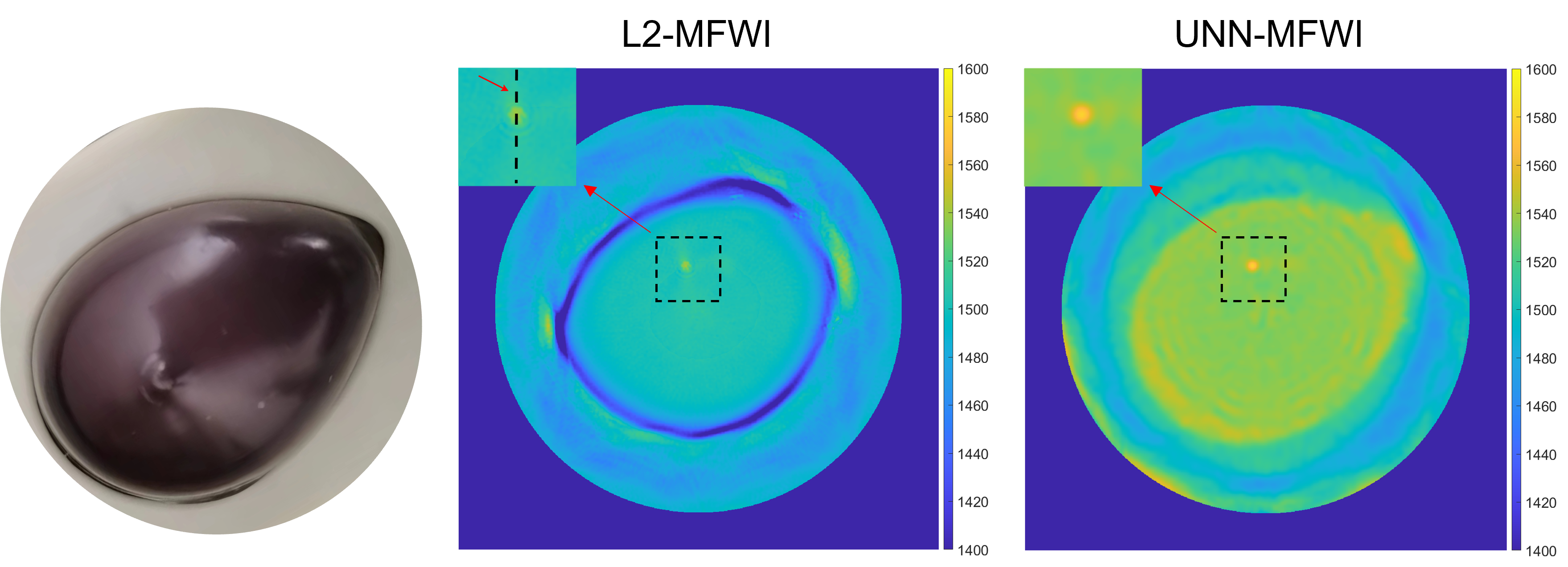}
\caption{Results of the phantom experiment on the breast. The left panel shows the phantom model, the middle panel shows the L2-MFWI result, and the right panel shows the UNN-MFWI result. The area indicated by the black dashed line in the rectangular box in the figure was used for quantitative analysis, and the results are shown in Fig. 7.}
\label{fig_5}
\end{figure*}

\section{Results}
In this section, we show the results of numerical simulation and phantom experiment of the breast. Then we present the comparison of the proposed method with the baseline method in terms of quantitative metrics and visualization.\par
\subsection{Numerical Simulation}
In the numerical simulation, Fig. 4 presents the numerical experimental results for different center frequencies. Fig. 4(a) depicts the reconstruction results of different methods at a frequency of 500K. It is obvious that the classical L2-FWI method performs poorly, with numerous artifacts present in the center. This suggests that the method converges to a local optimal solution. On the other hand, the L2-MFWI, UNN-FWI, and UNN-MFWI methods all yield good results, successfully reconstructing most of the breast details. A comparison between L2-MFWI and L2-FWI reveals that the adoption of the multi-band inversion strategy is highly effective. Furthermore, a comparison between L2-MFWI and the UNN-FWI method demonstrates that the UNN method accurately reconstructs the speed of sound image even without the use of the multi-band strategy. This affirms the effectiveness of the UNN framework for FWI, showing that it has a regularization effect and alleviates the ill-posed nature of the FWI problem. Moreover, when examining the details of the breast skin, it is evident that the reconstructed images generated by the UNN-FWI and UNN-MFWI methods are more accurate than those produced by L2-MFWI. Finally, comparing the UNN-FWI and UNN-MFWI methods reveals minimal differences in the reconstruction results, with only slightly higher quantitative metrics observed for UNN-MFWI. This indicates that the UNN method is less influenced by multi-band strategies compared to L2-FWI in the low-frequency case.\par
The results for the 800K center frequency can be observed in the Fig. 4(b). It is clear that the task of reconstructing sound speed using FWI becomes more non-convex and non-linear as the frequency increases. Among the four methods, only the UNN-MFWI method achieves accurate reconstruction, while the other three methods generate significant artifacts in the central region. In detail among the three failed methods, the UNN-FWI method works slightly better than the other two classical approaches in some glandular regions from a visual perspective, and the quantitative metrics also show that UNN-FWI is slightly better. These findings suggest that the UNN-MFWI approach exhibits greater robustness than UNN-FWI in high-frequency cases. Furthermore, incorporating a multi-band strategy into the UNN framework improves the ill-posed problem even further. This also demonstrates the adaptability of the UNN architecture to other commonly used FWI methods for addressing the cycle-skipping problem. Combining UNN with these methods has the potential to yield better outcomes.

\begin{table}[ht]
\centering
    \caption{Quantitative Measures of Numerical Experiments By Different Methods.}
    \begin{tabular}{ccccccccccccc}
    \toprule
    & \multicolumn{6}{c}{Breast model} \\
    & \multicolumn{3}{c}{500K} & \multicolumn{3}{c}{800K} \\ 
    \cmidrule(lr){2-4}\cmidrule(lr){5-7} 
      Method        & SSIM   & PSNR   & RMSE  & SSIM  & PSNR  & RMSE \\
      \midrule
      L2-FWI       & 0.7984 & 29.19 &  0.0413   &  0.6259  & 26.73   &  0.0600 \\
      L2-MFWI      & 0.9519 & 41.28 & 0.0048    &  0.7387  &  21.94  &  0.0925 \\
      UNN-FWI          & 0.9974 & 50.36 & 0.0035     &  0.9616  &   28.25 &   0.0493 \\
      UNN-MFWI         & \textbf{0.9981} & \textbf{53.30} & \textbf{0.0026}      &  \textbf{0.9976}  &  \textbf{52.33}  & \textbf{0.0028} \\
      \bottomrule
\end{tabular}
\end{table}

\subsection{Phantom Experiment}
Fig.5 shows the results of the phantom experiment. The comparison in this experiment was limited to the UNN-MFWI and MFWI methods for simplicity. Fig. 5 demonstrates that the FWI method, both with and without the UNN module, effectively reconstructs most of the region of the breast phantom. In particular, the high sound speed scattering point is visible in both methods. However, we can find that the result of the L2-MFWI method has two major flaws: first, it generates a distinct band of low sound speed errors at the breast-water interface; second, the outer contour of the reconstructed breast is irregular. The reconstruction quality is poor, particularly in the upper right corner of the breast. The UNN-MFWI result mitigates both of these drawbacks. It has the same contour as the breast phantom and, besides that, it eliminates the low sound speed band artifacts. The reason for the large error in the contour of the FWI reconstruction results is hypothesized to be that the breast phantom is not a cylindrical shape in three dimensions, and thus off-axis reflections and transmissions may occur during the process of performing the two-dimensional scanning, the phenomenon that is particularly severe with the large incidence angle at the upper right corner. This results in the observed signal containing a significant systematic error due to the two-dimensional assumption. In addition, the off-axis phenomenon leads to a lower energy of the received signal, making the observed signal have a low signal-to-noise ratio. From the final reconstruction results, the robustness of the UNN-MFWI method to noise and systematic errors is better than that of the L2-MFWI method. The UNN module can enhance the performance of the FWI sound speed reconstruction algorithm under low-quality observed signals.\par
From the quantitative analysis, we chose a profile at y = 109.1 mm to measure the error between the reconstruction results and the standard ground truth, as shown in Fig. 6. From the results, it can be seen that the L2-MFWI result has a significant underestimation of the background and scattering point sound speed of the phantom, while the UNN-MFWI result is much closer to the real situation. The difference between the two methods is not significant in the performance of the scattering point, and both can reconstruct the location and size of the scattering point well.

\begin{figure}[!t]
\centering
\includegraphics[width=3in]{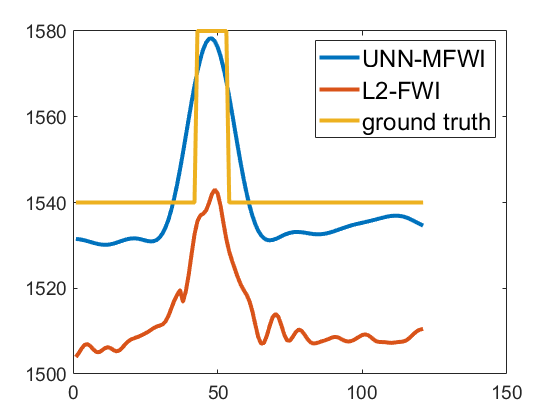}
\caption{Profiles through y = 109.1 mm for ground truth and reconstructed images of L2-MFWI and UNN-MFWI in the phantom experiment}
\label{fig_6}
\end{figure}

\section{Discussion}
Through a series of numerical simulations and the phantom experiment, this study demonstrates the effectiveness of the UNN architecture in improving high-resolution sound speed imaging in USCT. Impressive results were shown in this study. In comparison to the traditional FWI algorithm, the method proposed in this paper incorporates an untrained neural network into its iterative process. The network is utilized as a tool for generating sound speed maps and sparsely represents the distribution of sound speeds in human tissues using network parameters. Consequently, this approach achieves implicit regularization of a non-convex and strongly non-linear optimization problem.\par
The UNN-based method offers three advantages compared to the traditional FWI technique. 1. UNN shows strong performance against noise and systematic errors. In the presence of low-quality observed signals and large system modeling errors, the addition of the UNN module can substantially improve the algorithm's robustness. 2. In regions with significant variations in sound speed, such as the skin of the breast, the UNN-based method outperforms FWI in terms of reconstruction accuracy and does not exhibit significant artifacts. 3. In general, the UNN-FWI includes its built-in regularization effect, which helps prevent the sound speed reconstruction process from being trapped in local optimal solutions. This regularization effect mitigates the issue of cycle-skipping to some degree and consequently enhances the robustness of the method, particularly in high-frequency scenarios. Besides, the UNN framework can be integrated as a plug-and-play component into existing FWI variant algorithms. For instance, it can be applied to the multi-band strategy mentioned in this paper or other loss function-based variant algorithms. The latter integration only needs to modify the loss function instead of relying on the simple L2 norm. As far as we know, there is no fundamental conflict with other variants, ensuring a high level of compatibility with other methods. Moreover, embedding this module serves as an implicit regularization, potentially further improving the variant algorithm and facilitating convergence to the global optimal solution.\par
However, the method proposed in this paper still has some limitations. Firstly, there is a challenge concerning computational efficiency, as the current reconstruction process requires solving the forward problem of the wave equation repeatedly. This is known to be computationally expensive. In the experiments carried out in this paper, utilizing a center frequency of 500K, each forward computation requires 1.53 seconds. The total number of iterations is typically set at 5000, resulting in a cumulative forward computation time of approximately 7500 seconds. To tackle this issue, our future plan involves reducing the number of iterations through the implementation of an optimizer such as L-BFGS, known for its superlinear convergence properties. Additionally, we aim to accelerate the optimization process by incorporating a reference image loss term into the loss function, providing guidance for faster convergence. The second concern pertains to the issue of overfitting, wherein an excessive number of iterations can inadvertently lead the network to learn and incorporate the input noise distribution, leading to the presence of speckle noise in the reconstructed image. To address this, it becomes essential to identify an appropriate termination point, commonly referred to as the early-stopping strategy, which is currently determined by empirical knowledge. So, our future plan involves implementing an adaptive early-stopping strategy to eliminate the need for expert intervention.

\section{Conclusion}
In this paper, we propose a novel plug-and-play neural network module for the reconstruction of high-resolution sound speed images in USCT. Our method utilizes an untrained neural network as a generator. The input of the network is noise, and the output is a non-quantitative image that captures information about the distribution of sound speed. This image is then transformed into a quantitative sound speed distribution image using a transformation operator. The subsequent steps follow the concept of FWI, where the sound speed image is mapped to ultrasound signal data through a forward process. The resulting signal is compared with the observed signal, and the discrepancies are used to update the network parameters through a backward process. This iterative process aims to produce a sound speed image that closely represents the real situation. The proposed method acts as an implicit regularisation of the original FWI problem through the network structure. The superior performance of this method is demonstrated through numerical simulation and phantom experiments conducted on the breast. By incorporating this module, the sensitivity of FWI to initial values is mitigated, making it less prone to encounter cycle-skipping phenomenon. Besides, the proposed method is insensitive to noise and systematic errors, and has good performance near the strong sound speed contrast region. Additionally, because the module exclusively influences the generation of sound speed images without impacting the overall FWI process or its variations, it can be effortlessly integrated into existing algorithmic frameworks as a standalone plug-and-play module. Moreover, experimental evidence shows that the inclusion of this module further enhances the robustness of the better-performing variants of the FWI in high-frequency scenarios. Therefore, it is believed that the proposed method offers a feasible alternative to the current sound speed imaging in USCT.
\IEEEpubidadjcol


\bibliographystyle{IEEEtran}
\bibliography{IEEEabrv,ref}

\end{document}